\documentclass[12pt]{article}
\usepackage{epsf}

\newcommand{\bea}{$$\begin{array}{rcl}}
\newcommand{\eea}{\end{array}$$}

\def\vev#1{\left\langle #1\right\rangle}

\def\Im{\mathop{\mbox{Im}}}

\def\ee{$\varepsilon'/\varepsilon$~}

\title{Estimating $\varepsilon'/\varepsilon$ in the Standard Model} 

\author{S. Bertolini
\\
INFN, Sezione di Trieste\\
Scuola Internazionale Superiore di Studi Avanzati\\
via Beirut 4, I-34013 Trieste, Italy.}

\date{June 22, 1999}

\begin{document}

\maketitle

\begin{abstract}
I discuss the comparison of the current theoretical calculations
of \ee with the experimental data. Lacking reliable 
``first principle'' calculations, 
phenomenological approaches may help in understanding 
correlations among different contributions and available experimental data.
In particular, in the chiral quark model approach the same dynamics which
underlies the $\Delta I = 1/2$ selection rule in kaon decays 
appears to enhance the $K\to\pi\pi$ matrix element of
the $Q_6$ gluonic penguin, thus driving \ee in the range of the
recent experimental measurements.
\end{abstract} 

\vspace{1cm}



The results announced by the KTeV Collaboration last February and 
by the NA48 Collaboration at this conference~\cite{Sozzi}
(albeit preliminary) have marked a great experimental achievement,
establishing 35 years after the discovery of CP violation
in the neutral kaon system
the existence of a much smaller violation acting directly in the
decays. 

While the Standard Model (SM) of strong and electroweak interactions
provides an economical and elegant understanding of the presence
of indirect ($\varepsilon$) and direct ($\varepsilon'$) 
CP violation in term of a single phase,
the detailed calculation of the size of these effects 
implies mastering strong interactions at a scale 
where perturbative methods break down. In addition, CP violation
in $K\to\pi\pi$ decays
is the result of a destructive interference between
two sets of contributions (for a suggestive picture of 
the gluonic and electroweak penguin diagrams
see the talk by Buras at this conference~\cite{Buras}),
thus potentially inflating up to an order of magnitude the uncertainties
on the individual hadronic matrix elements of the effective 
four-quark operators.

\begin{figure}
\epsfxsize=9cm
\centerline{\epsfbox{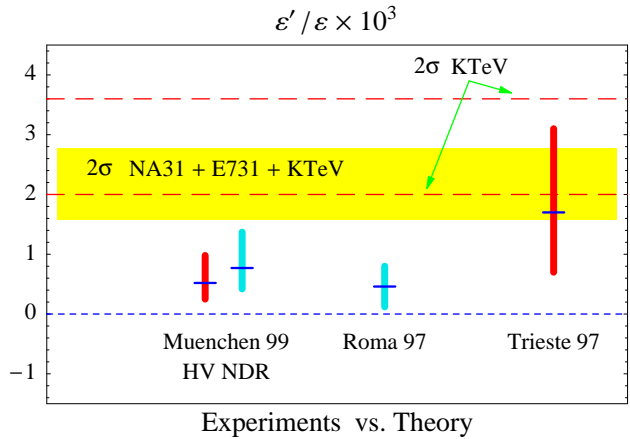}}
\caption{The recent KTeV result ($2\sigma$) is shown by
the area enclosed by the long-dashed lines. 
The combined $2\sigma$ average of the KTeV,
NA31 and E731 results is shown by the gray band.  
The predicted M\"unchen, Roma and Trieste theoretical ranges for \ee are shown 
by the vertical bars with their central values.}
\label{fig3new}
\end{figure}

In Fig.~\ref{fig3new}, taken from Ref.~\cite{BEFrev}, the comparison 
of the theoretical 
predictions and the experimental results available before 
the Kaon 99 conference is summarized.
The gray horizontal band shows
the two-sigma experimental range obtained averaging the recent KTeV result
with the older NA31 and E731 data, corresponding to 
$\varepsilon'/\varepsilon = (21.8 \pm 3) \times 10^{-4}$. 
The vertical lines show the ranges of the most
recent published theoretical predictions, identified with the cities where
most of the group members reside.
The figure does not include two new results announced at this conference:
on the experimental side, the first NA48 measurement~\cite{Sozzi} and, on the
theoretical side, the new
prediction based on the $1/N$ expansion~\cite{Hambye}, 
which I will refer to in the 
following as the Dortmund group estimate. 
The inclusion of the NA48 result 
$\varepsilon'/\varepsilon = (18.5 \pm 7.3) \times 10^{-4}$ 
lowers the experimental average shown in
Fig.~\ref{fig3new} by about 4\%.

Looking at Fig.~\ref{fig3new} two comments are in order. 
On the one hand, we should appreciate the fact
that within the uncertainties of the theoretical calculations, there is indeed
an overall agreement among the different predictions. 
All of them agree on the presence of a non-vanishing positive effect
in the SM.
On the other hand,
the central values of the M\"unchen (phenomenological $1/N$)
and Rome (lattice) calculations 
are by a factor 3 to 5 lower than the averaged experimental central value.

In spite of the complexity of the calculations, I would like to emphasize
that the difference between the predictions of the two estimates above 
and that of the Trieste group, 
based on the Chiral Quark Model ($\chi$QM)~\cite{Weinberg-GeorgiManohar},
is mainly due to the different size of the 
hadronic matrix element of the gluonic penguin $Q_6$.
In addition, I will show that the enhancement
of the $Q_6$ matrix element in the $\chi$QM approach
can be simply understood in terms
of chiral dynamics and, in this respect, it is related
to the phenomenological embedding of the $\Delta I= 1/2$ selection rule.

The $\Delta I = 1/2$ selection rule in $K\to\pi\pi$ decays is known by 
some 40 years~\cite{Pais-Gell-Mann} and it states the fact that kaons
are 400 times more likely to decay in the isospin zero two-pion state
than in the isospin two component. This rule is not justified by any
symmetry consideration and, although it is common understanding 
that its explanation must be rooted in the dynamics of strong interactions,
there is no up to date derivation of this effect from first principle QCD. 

As summarized by Martinelli at this conference~\cite{Martinelli}
lattice cannot provide us at present with reliable calculations
of the $I=0$ penguin operators relevant to \ee, as well as of the 
$I=0$ components of the hadronic matrix elements of the tree-level
current-current operators (penguin contractions), which are relevant for the
$\Delta I = 1/2$ selection rule.

\begin{figure}
\epsfxsize=9cm
\centerline{\epsfbox{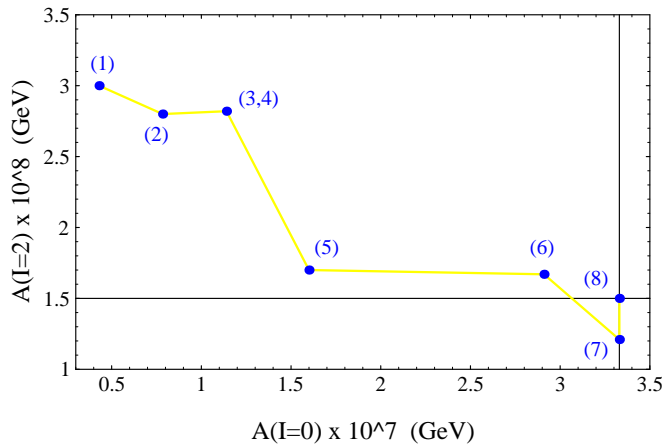}}
\caption{Anatomy of the $\Delta I = 1/2$ rule in the 
$\chi$QM~\cite{BEFdel-eps}. See the text for explanations.
The cross-hairs indicate the experimental point.}
\label{road}
\end{figure}

In the M\"unich approach the $\Delta I = 1/2$ rule
is used in order to determine phenomenologically the  
matrix elements of $Q_{1,2}$ and, 
via operatorial relations, some of the matrix elements of the
left-handed penguins. Unfortunately, the approach does not allow
for a phenomenological determination of the matrix elements of the penguin
operators which are most relevant for \ee, namely the gluonic penguin $Q_6$
and the electroweak penguin $Q_8$. 
Values in the ballpark of the leading $1/N$ estimate are assumed
for these matrix elements,
taking also into account that all present approaches
show a suppression of $\vev{Q_8}$ with respect to 
its vacuum saturation approximation (VSA).

In the $\chi$QM approach, the hadronic matrix elements
can be computed as an expansion in momenta
in terms of three parameters:
the constituent quark mass, the quark condensate
and the gluon condensate.
The Trieste group has computed the $K\to\pi\pi$ matrix elements
of the $\Delta S =1,2$ effective lagrangian up to $O(p^4/N)$ in the
chiral and $1/N$ expansions~\cite{BEFdel-eps}. 

Hadronic matrix elements and short distance Wilson coefficients are matched
at a scale of $0.8$ GeV as a reasonable
compromise between the ranges of validity 
of perturbation theory and chiral lagrangian.
By requiring the $\Delta I = 1/2$ rule to be reproduced
within a 20\% uncertainty one obtains a phenomenological
determination of the three basic parameters of the model.
This step is needed
in order to make the model predictive, since there is no a-priori argument
for the consistency of the matching procedure. As a matter of
fact, all computed observables turn
out to be very weakly scale dependent in a few hundred MeV range around the 
matching scale.

\begin{figure}
\epsfxsize=9cm
\centerline{\epsfbox{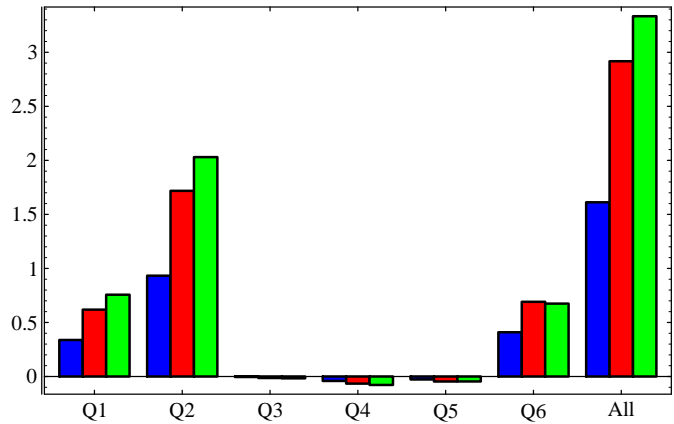}}
\caption{Anatomy of the $A(K^0\to\pi\pi)_{I=0}$ amplitude ($A_0$)
in units of $10^{-7}$ GeV
for central values of the $\chi$QM input parameters~\cite{BEFdel-eps}:
$O(p^2)$ calculation (black), with minimally subtracted
chiral loops (half-tone), complete $O(p^4)$ result (light gray).}
\label{chart0}
\end{figure}

Fig.~\ref{road} shows an anatomy of the (model dependent) contributions
which lead in the Trieste approach to reproducing
the $\Delta I = 1/2$ selection rule.

Point (1) represents the result obtained by neglecting QCD
and taking the factorized matrix element for the
tree-level operator $Q_2$, which is the only one present. 
The ratio $A_0/A_2$ is found equal
to $\sqrt{2}$: a long way to the experimental point (8).
Step (2) includes the effects of perturbative QCD renormalization
on the operators $Q_{1,2}~$\cite{Gaillard-etc}. 
Step (3) shows the effect of including the gluonic 
penguin operators~\cite{VSZ-GWise-CFGeorgi}. 
Electroweak penguins~\cite{Lusignoli-etc} are numerically
negligeable for the CP conserving amplitudes and 
are responsible for the very small shift in the $A_2$ direction.
Therefore, perturbative QCD and factorization lead us from (1) to (4).

Non-factorizable gluon-condensate corrections,
a crucial model dependent effect, enter at the leading order
in the chiral expansion leading to a substantial
reduction of the $A_2$ amplitude (5), 
as first observed by Pich and de Rafael~\cite{Pich-deRafael}. 
Moving the analysis to $O(p^4)$
the chiral loop corrections, computed on the LO chiral
lagrangian via dimensional regularization and minimal subtraction, 
lead us from (5) to (6), while 
the corresponding $O(p^4)$ tree level counterterms
calculated in the $\chi$QM lead to the point (7).
Finally, step (8) represents the inclusion of $\pi$-$\eta$-$\eta'$ 
isospin breaking effects~\cite{Ometapeta}. 

{\footnotesize
\begin{table}
\begin{center}
\begin{tabular}{c|| c| c| c}
$B_i$ & M\"unchen 99 & Roma 97 & Trieste 97 \\
 &      $\mu = 1.3$ GeV & $\mu = 2.0$ GeV &  $\mu = 0.8$ GeV\\
\hline
 $B_1^{(0)}$& 13~$(\dag)$ &  $-$ &9.5  \\
 $B_2^{(0)}$& 6.1~$(\dag)$  & $-$  & 2.9  \\
 $B_1^{(2)} = B_2^{(2)}$& 0.48~$(\dag)$ &  $-$ & 0.41 \\
 {$B_3$} & {1~(*)}  & {1~(*)} & {$-2.3$} \\
 {$B_4$} & {5.2~(*\ \dag)} & {$1\div 6$~(*)} & {1.9}\\
 {$B_5 \simeq B_6$}& { $1.0\pm 0.3$~(*)} &
 { $1.0\pm 0.2$} & {  $1.6\pm 0.3$} \\
 $B_7^{(0)} \simeq B_8^{(0)}$&1~(*)   & 1~(*) & $2.5\pm 0.1$  \\
 $B_9^{(0)}$& 7.0~(*\ \dag)  &1~(*)  & 3.6 \\
 $B_{10}^{(0)}$& 7.5~(*\ \dag)  &1~(*)  & 4.4 \\
 $B_7^{(2)}$&1~(*)  &$ 0.6\pm 0.1$  &  $0.92\pm 0.02$ \\
 {$B_8^{(2)}$}& { $0.8\pm 0.15$~(*)}  &
 {$ 0.8\pm 0.15$}  & { $0.92\pm 0.02$} \\
 $B_9^{(2)}$& 0.48~$(\dag)$   & $0.62\pm 0.10$ & 0.41  \\
 $B_{10}^{(2)}$&0.48~$(\dag)$   & 1~(*) & 0.41 \\
\hline
 {$\hat B_K$}& {$ 0.80\pm 0.15$}   &
 {$ 0.75\pm 0.15$}  & {$ 1.1\pm 0.2$}  \\
\hline
\end{tabular}
\end{center}
\caption{Summary of $B$ factors.
Legenda:  (*) educated guess, $(\dag)$ derived from the $\Delta I = 1/2$ rule.
In the Trieste calculation the $\Delta I = 1/2$ rule is used to constrain the
three basic model parameters in terms of which all matrix elements 
are computed.}
\label{Bfactors}
\end{table}
}

This model dependent anatomy 
shows the relevance of non-factorizable contributions
and higher-order chiral corrections. The suggestion that
chiral dynamics may be relevant to
the understanding of the $\Delta I = 1/2$ selection rule goes back to the
work of Bardeen, Buras and Gerard~\cite{BBG,Bardeen} in the $1/N$
framework using a cutoff regularization. This approach
has been recently revived and improved 
by the Dortmund group, with a particular attention
to the matching procedure~\cite{Hambye,Bardeen}. 
A pattern similar to that shown
in Fig.~\ref{road} for the chiral loop corrections to $A_0$ and $A_2$
was previously obtained in a NLO chiral
lagrangian analysis, using dimensional regularization, by 
Missimer, Kambor and Wyler~\cite{MKWyler}.

The $\chi$QM approach allows us to further investigate the relevance
of chiral corrections for each of the effective quark operators
of the $\Delta S = 1$ lagrangian.  
Fig.~\ref{chart0} shows the contributions to the CP conserving
amplitude $A_0$ of the relevant operators, providing us
with a finer (model dependent) anatomy of the NLO chiral corrections.
From Fig.~\ref{chart0} we notice that, because of the chiral loop enhancement,
the $Q_6$ contribution to $A_0$
is about 20\% of the total amplitude. As we shall see,
the $O(p^4)$ enhancement of the $Q_6$ matrix element is what drives \ee 
in the $\chi$QM to the $10^{-3}$ ballpark.

A commonly used way of comparing the estimates of hadronic matrix elements
in different approaches is via the so-called $B$ factors which
represent the ratio of the model matrix elements to the corresponding VSA
values. However, care must be taken in the comparison of
different models due to the scale
dependence of the $B$'s and the values used by different groups
for the parameters that
enter the VSA expressions. Table \ref{Bfactors} reports the
$B$ factors used for the predictions shown in Fig.~\ref{fig3new}.
 
\begin{figure}
\epsfxsize=9cm
\centerline{\epsfbox{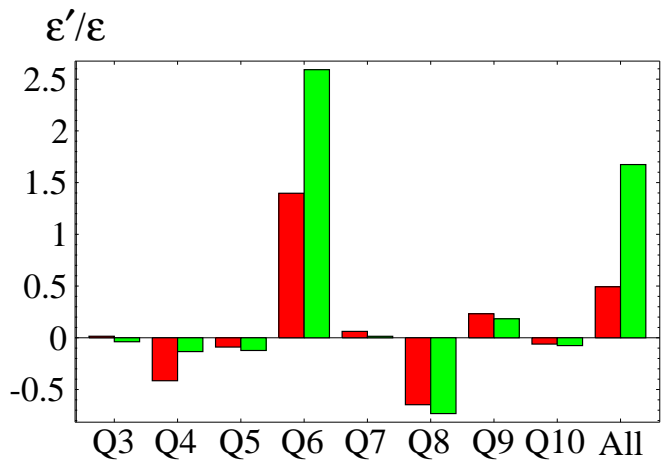}}
\caption{Predicting $\varepsilon'/\varepsilon$:
a (Penguin) Comparative Anatomy of the M\"unchen (dark gray) and 
Trieste (light gray) results (in units of $10^{-3}$).}
\label{istobuqm}
\end{figure}

An alternative pictorial and synthetic
way of analyzing different outcomes for \ee
is shown in Fig.~\ref{istobuqm}, where a ``comparative
anatomy'' of the Trieste and M\"unchen estimates is presented. 

From the inspection of the various contributions it is apparent that
the final difference on the central value of \ee is almost entirely
due to the difference in the $Q_6$ component. In second order, a larger
(negative) contribution of the $Q_4$ penguin in the M\"unchen calculation
goes into the direction of making \ee smaller.

The difference in the $Q_4$ contribution is easily understood. In the
M\"unchen estimate the $Q_4$ matrix element is obtained using the operatorial
relation $\vev{Q_4} = \vev{Q_2}_0 - \vev{Q_1}_0 + \vev{Q_3}$,
together with the knowledge acquired on $\vev{Q_{1,2}}_0$ from fitting 
the $\Delta I = 1/2$ selection rule at the charm scale.

As a matter of fact, 
the phenomenological fit of $\Delta I = 1/2$ rule requires 
a large value of $\vev{Q_2}_0 - \vev{Q_1}_0$ (which deviates by up to an
order of magnitude from the naive VSA estimate). The assumption that
$\vev{Q_3}$ is given by its VSA value leads,
in the M\"unchen analysis, to a large value of $\vev{Q_6}$: about 5 times
larger than its VSA value. 
On the other hand, in the $\chi$QM calculation $\vev{Q_3}$ turns out to
have a sign opposite to its VSA expression, in such a way that
a smaller value for $Q_4$ is obtained.
A lattice calculation of all gluonic penguins is definitely needed
to disentangle such patterns. 

\begin{figure}
\epsfxsize=9cm
\centerline{\epsfbox{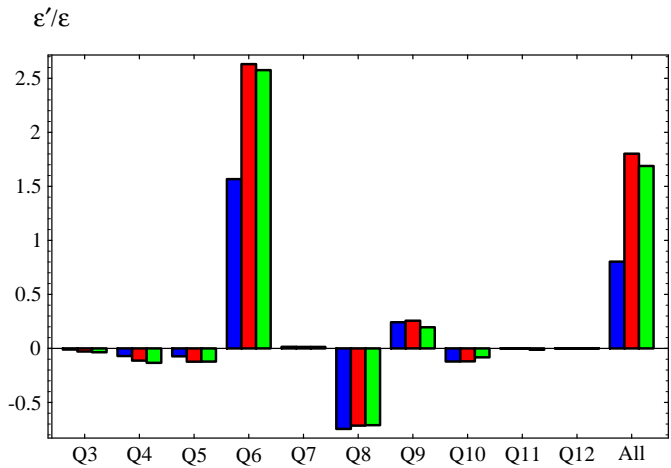}}
\caption{Anatomy of \ee (in units of $10^{-3}$) within the $\chi$QM 
approach~\cite{BEFdel-eps}. 
In black the LO results (which includes the non-factorizable gluonic
corrections), in half-tone the effect of the inclusion of chiral-loop 
corrections and in light gray the complete $O(p^4)$ estimate.}
\label{charteps}
\end{figure}

At any rate, the main difference between the \ee central values
obtained in the Trieste and M\"unchen calculations
rests in the $Q_6$ matrix element. The nature of the difference 
is apparent in Fig.~\ref{charteps} where the various penguin
contributions to \ee in the Trieste analysis
are further separated in LO (dark histograms) and NLO
components---chiral loops (gray histograms) and $O(p^4)$ tree level
counterterms (dark histograms).

It is clear that chiral loop dynamics plays a subleading role in the 
electroweak penguin sector ($Q_{8-10}$) while enhancing by 60\% the $Q_6$
matrix element. 
At $O(p^2)$ the $\chi$QM prediction
for \ee would just overlap with the M\"unchen estimate once the small
effect of the $Q_4$ operator is taken into account. 

The $\chi$QM analysis shows that the same dynamics that is relevant to the
reproduction of the CP conserving $A_0$ amplitude 
(Fig.~\ref{chart0}) is at work
also in the CP violating sector (gluonic penguins).

In order to ascertain
whether the model features represent real QCD effects one should
wait for future improvements in lattice calculations ~\cite{Blum}.
On the other hand,
indications for such a dynamics arise from current $1/N$  
calculations~\cite{Hambye,Bijnens}. 

The idea of a connection between the $\Delta I = 1/2$ selection
rule and \ee is certainly not new~\cite{VSZacharov,Gilman-Wise},
although at the GeV scale, where one can trust perturbative
QCD, penguins are far from providing
the dominant contribution to the CP conserving amplitudes.

Before concluding, I like to make a comment on the role of
the strange quark mass in the $\chi$QM calculation of \ee:
in such an approach the basic parameter that enters the relevant
penguin matrix elements is the quark condensate and 
the explicit dependence
on $m_s$ appears at the NLO in the chiral expansion.
Varying the central value of $\bar m_s(m_c)$ from 150 MeV to 130 MeV
affects $\vev{Q_6}$ and  $\vev{Q_8}$ at the few percent level.

A more sensitive quantity is $\hat B_K$, which parametrize the
$\bar K- K$ matrix element. This parameter, which equals unity in the VSA
turns out to be quite sensitive to $SU(3)$ breaking effects.
Taking $\bar m_s(m_c)=130\pm 20$ MeV, 
$\Lambda_{\rm QCD}^{(4)}=340\pm 40$ MeV
and varying all relevant paramaters, the updated $\chi$QM result is:
$$
\hat B_K = 1.0 \pm 0.2\ ,
$$
to be compared with the value used in the 1997 analysis (Table \ref{Bfactors}).

This increases the previous determination of $\Im \lambda_t$ by roughly
10\% and correspondingly \ee (an updated analysis  of \ee in the $\chi$QM
with gaussian treatment of experimental inputs is in progress).
 
I conclude by summarizing the relevant remarks:

\begin{itemize}
\item
Phenomenological approaches which embed the
$\Delta I = 1/2$ selection rule in $K\to\pi\pi$ decays,
generally agree with present lattice calculations
in the pattern and size of the $I=2$ components
of the $\Delta S = 1$ hadronic matrix elements.

\item
Concerning the $I=0$ matrix elements, where lattice
calculations suffer from large sistematic uncertainties, the 
$\Delta I = 1/2$ rule forces upon us large deviations from
the naive VSA (see Table \ref{Bfactors}).

\item
In the Chiral Quark Model calculation, the fit of the CP conserving
$K\to\pi\pi$ amplitudes, which determines the three 
basic parameters of the model, feeds down to the penguin sectors
showing a substancial enhancement of the $Q_6$ matrix element, such that
$B_6/B_8^{(2)} \approx 2$. This is what drives the \ee prediction
in the $10^{-3}$ ballpark.

\item
Up to {40\%} of the present uncertainty in the \ee prediction arises
from the uncertainty in the CKM elements $\Im (V_{ts}^*V_{td})$ 
which is presently controlled by the $\Delta S =2$ parameter $B_K$.
A better determination of the unitarity triangle from B-physics
is expected from the B-factories and hadronic colliders.
From K-physics, the rare decay $K_L\to\pi^0\nu\bar\nu$ 
gives the cleanest ``theoretical''
determination of $\Im\lambda_t$~\cite{Buras}.

\item
In spite of clever and interesting new-physics explanations for a large 
\ee~\cite{Hall,Murayama,Isidori,Benatti},
it is premature to interpret the present theoretical-experimental
``disagreement'' as a signal of physics beyond the SM. 
Ungauged systematic uncertainties affect presently all theoretical 
estimates. Not to forget the long-standing puzzle of the $\Delta I = 1/2$
rule: perhaps an ``anomalously'' large \ee 
($B_6/B_8^{(2)} \approx 2$) is just the CP violating projection 
of $A_0/A_2 \approx 20$.
\end{itemize}

My appreciation goes to the Organizers of the Kaon 99 Conference for
assembling such a stimulating scientific programme and for the
efficient logistic organization. 
Finally, I thank J.O. Eeg and M. Fabbrichesi for a most enjoyable and fruitful 
collaboration.


\begin{thebibliography}{99}

\bibitem{Sozzi} M. Sozzi, these proceedings.

\bibitem{Buras} A.J. Buras, these proceedings.

\bibitem{BEFrev} S. Bertolini, J.O. Eeg and M. Fabbrichesi, hep-ph/9802405,
                    to appear in Reviews of Modern Physics.

\bibitem{Hambye} T. Hambye, these proceedings.

\bibitem{Weinberg-GeorgiManohar} S. Weinberg, Physica A96 (1979) 327;
               A. Manhoar and H. Georgi, Nucl. Phys. B234 (1984) 189.

\bibitem{Pais-Gell-Mann} M. Gell-Mann and A. Pais, 
                         Proc. Glasgow Conf. p.342, 1955.

\bibitem{Martinelli} G. Martinelli, these proceedings. 

\bibitem{BEFdel-eps} S. Bertolini, J.O. Eeg, M. Fabbrichesi and E.I. Lashin, 
                 Nucl. Phys. B514 (1998) 63; 
        Nucl. Phys. B514 (1998) 93. 

\bibitem{Gaillard-etc} M.K. Gaillard and B. Lee, 
          Phys. Rev. Lett. 33 (1974) 108; 
          G. Altarelli and L. Maiani, Phys. Lett. B52 (1974) 413. 
 
\bibitem{VSZ-GWise-CFGeorgi} M.A. Shifman, A.I. Vainshtein and V.I. Zacharov,
JETP Lett. 22 (1975) 55; Nucl. Phys. B120 (1977) 316; 
Sov. Phys. JETP 45 (1977) 670;
J. Ellis, M.K. Gaillard, D.V. Nanopoulos and S. Rudaz,
Nucl. Phys. B131 (1977) 285, (E) B132 (1978) 541;
F.J. Gilman and M.B. Wise, Phys. Lett. B83 (1979) 83;
Phys. Rev. D20 (1979) 2392; R.S. Chivukula, J.M. Flynn and H. Georgi,
Phys. Lett. B171 (1986) 453.

\bibitem{Lusignoli-etc} J. Bijnens and M.B. Wise, Phys. Lett. B137 (1984) 245; 
A.J. Buras and J.M. Gerard, Phys. Lett. B192 (1987) 156; 
M. Lusignoli, Nucl. Phys. B325 (1989) 33; J.M. Flynn and L. Randall,
Phys. Lett. B224 (1989) 221.   

\bibitem{Pich-deRafael} A. Pich and E. de Rafael, Nucl. Phys. B358 (1991) 311.

\bibitem{Ometapeta} B. Holstein, Phys. Rev. D20 (1979) 1187; 
J.F. Donoghue, E. Golowich, 
B.R. Holstein and J. Trampetic, Phys. Lett. B179 (1986) 361; 
A.J. Buras and J.M. Gerard, in ref. \cite{Lusignoli-etc}.

\bibitem{BBG} W.A. Bardeen, A.J. Buras and J.M. Gerard, Nucl. Phys. B293
(1987) 787.

\bibitem{Bardeen} W.A. Bardeen, these proceedings. 

\bibitem{MKWyler} J. Kambor, J. Missimer and D. Wyler, Nucl. Phys. B346
(1990) 17; Phys. Lett. B261 (1991) 496. 

\bibitem{Blum} T. Blum, these proceedings. 

\bibitem{Bijnens} J. Bijnens, these proceedings.

\bibitem{VSZacharov} M.A. Shifman, A.I. Vainshtein and V.I. Zacharov, 
in ref. \cite{VSZ-GWise-CFGeorgi}.

\bibitem{Gilman-Wise} F.J. Gilman and M.B. Wise,
in ref. \cite{VSZ-GWise-CFGeorgi}.

\bibitem{Hall} L. Hall, these proceedings. 

\bibitem{Murayama} H. Murayama, these proceedings.

\bibitem{Isidori} G. Isidori, these proceedings.

\bibitem{Benatti} F. Benatti and R. Floreanini, hep-ph/9906272.

\end{thebibliography}
\end{document}